\documentclass[a4paper,11pt]{article}
\usepackage{pos}

\newcommand*{\no}{\noindent}
\newcommand*{\bea}{\begin{eqnarray}}
\newcommand*{\eea}{\end{eqnarray}}
\newcommand*{\be}{\begin{equation}}
\newcommand*{\ee}{\end{equation}}
\newcommand*{\pd}{\partial}

\newcommand*{\pref}[1]{(\ref{#1})}

\newcommand*{\mn}{{\mu\nu}}

\newcommand*{\str}{\mathrm{str}}

\newcommand*{\nn}{\nonumber}

\newcommand*{\tl}{\mathrm{tl}}

\newcommand{\bma}{\begin{pmatrix}}
\newcommand{\ema}{\end{pmatrix}}

\newcommand*{\op}{{{\cal O}}}
\newcommand*{\la}{\left\langle}
\newcommand*{\ra}{\right\rangle}

\title{Towards testing the Fr\"ohlich-Morchio-Strocchi mechanism in quantum gravity}
\author*[a]{Axel Maas}

\affiliation[a]{Institute of Physics, NAWI Graz, University of Graz, Universit\"atsplatz 5, A-8010 Graz, Austria}

\emailAdd{axel.maas@uni-graz.at}

\abstract{Physics in (canonical) quantum gravity needs to be manifestly diffeomorphism-invariant. Consequently, physical observables need to be formulated in terms of manifestly diffeomorphism-invariant operators, which are necessarily composite. This makes an evaluation in general involved, even if the concrete implementation of quantum gravity should be treatable (semi-)perturbatively in general. A mechanism developed by Fröhlich, Morchio and Strocchi for flat-space-time quantum field theory may also be applicable in this case. It may be possible to test this mechanism with methods like dynamical triangulation.}

\FullConference{%
  41st International Conference on High Energy physics - ICHEP2022\\
  6-13 July, 2022\\
  Bologna, Italy
}

\begin{document}
\maketitle

\section{Introduction}

Observables should not depend on the choice of coordinate system. Thus, they need to be invariant under changes of the coordinate system. In quantum gravity, this translates into invariance under diffeomorphisms. This excludes quantities like the metric from being actually physical. However, the simplest invariant quantities are then scalar curvature invariants. From the point of view of a quantum field theory (QFT), these are involved composite operators. In fact, they are akin to those operators describing bound states in flat-space-time QFT.

As such, they are notoriously difficult to evaluate, especially outside lattice-type methods. It is here, were an approach developed by Fröhlich, Morchio and Strocchi (FMS) \cite{Frohlich:1980gj,Maas:2017wzi} for flat-space quantum gauge field theory may be helpful \cite{Maas:2019eux}. Its basic idea is relatively straightforward. It requires that the system in question is dominated by certain field values around which it only fluctuates weakly on average. However, in contrast to semi-classical approaches it is an exact treatment of the theory. Especially, it starts out from manifest, non-perturbatively diffeomorphism invariant correlation functions.

To implement it, the quantized theory is fixed to a gauge which is satisfied by the dominating field configuration. After that, and not before as in background-field methods, a split into the dominating part and fluctuation fields is performed. This split is performed both in the action and also inside the invariant correlation functions. The latter expands the correlation function into a sum of correlation functions. These can then be calculated within any suitable approximation scheme compatible with the smallness of fluctuations, usually perturbation theory in the splitted fields, which can be considered a small-field expansion. This approach yields very good results, also in comparison to numerical simulations and experiment, in suitable flat-space-time quantum gauge theories, in particular electroweak physics \cite{Maas:2017wzi}. In fact, the Brout-Englert-Higgs effect is part of this procedure.

Astronomical observations and microscopical experiments are compatible with the dominance of a classical metric of quantum gravity at distance scales much larger than the Planck scale. Thus, this is indeed similar to electroweak physics and its dominance by the Higgs vacuum expectation value. Therefore, the same approach may work as well in quantum gravity.

\section{The formalism}

The FMS approach is fairly independent of the actual theory. For the sake of concreteness, the following will use Einstein-Hilbert gravity with the metric as integration variable in a path integral, under the assumption that it enjoys asymptotic safety. The latter is supported by both renormalization group studies \cite{Reuter:2019byg,Bonanno:2020bil} as well as dynamical triangulation calculations \cite{Loll:2019rdj,Ambjorn:2022naa}. Of course, it may still not be the right theory, but as long as the ultimate theory is a quantum gauge theory of diffeomorphisms, and possibly torsion \cite{Hehl:1976kj}, the following will require only minor modifications \cite{Maas:2019eux}.

The starting point is a field configuration $g^c$ of the metric, which dominates the system. This is, of course, empirical input. While it appears tempting to use a Friedman-Robertson-Walker space-time as a good approximation to the history of our universe, such a choice has a severe limitation \cite{Maas:2022lxv}. Because it has a preferred event, the big bang, it breaks explicitly various symmetries. This is, like in a scattering process, a consequence of the initial conditions. And just like when investigating scattering processes in the FMS approach \cite{Maas:2017wzi}, such a feature should be moved into the correlation functions \cite{Maas:2022lxv}. As a consequence, $g^c$ should be a solution, which does not have preferred events. Assuming the existence of a cosmological constant, this leaves in Einstein-Hilbert gravity for now only Minkowski space-time and (anti)de Sitter space-time as choices for $g^c$. This is supported by dynamical triangulation calculations, which indicate that in absence of matter on average the geometry is indeed de Sitter \cite{Loll:2019rdj,Ambjorn:2022naa}.

The next step is to choose a gauge for the full metric, and quantize the theory. The situation is here aggravated in comparison to flat-space-time QFT, as choices are usually non-linear in the metric. Aside from this, the choice is dictated by technical convenience only. A possible choice \cite{Maas:2022lxv} is Haywood gauge, $\pd^\mu g_\mn=0$, as it is satisfied by the above mentioned choices for $g^c$. The theory is then quantized in terms of this gauge, where no reference is yet made to $g^c$. This is different to a background field approach. After this, the metric is split into $g^c$ and a fluctuation metric $\gamma$ \cite{Maas:2019eux}. As in the following a perturbative approach is pursued, $\gamma$ can be assumed to be small compared to $g^c$, and thus a linear split is sufficient \cite{Maas:2022lxv},
\be
g=g^c+\gamma\label{split}.
\ee
\no However, in contrast to $g$ and $g^c$, $\gamma$ is not a metric, and raising and lowering of indices is still performed with $g$. But the inverse of $\gamma$ obeys a Dyson equation \cite{Maas:2022lxv}
\be
\gamma^{-1}=-(g^c)^{-1}\gamma(g^c+\gamma)^{-1}\label{dyson}.
\ee
\no This equation can be solved by iteration to a consistent order in an expansion in $\gamma$ in expressions involving $g^{-1}$. However, this implies that there is now an infinite number of tree-level terms. Though if the theory is asymptotically safe, and thus renormalizable, these will not have independent couplings or counter terms.

So far, this is essentially the same procedure as for the Brout-Englert-Higgs effect, with $g^c$ taking over the role of the Higgs vacuum expectation value (vev). Thus, in the same sense the existence of a dominating field configuration $g^c$ breaks diffeomorphism symmetry as does the Higgs vev electroweak symmetry.

The core insight of the FMS approach is now to recognize that this breaking is just an artifact of the gauge fixing and splitting \cite{Frohlich:1980gj,Frohlich:1981yi,Maas:2017wzi}. For non-perturbatively diffeomorphism-invariant correlation functions, and especially observables, the symmetry is perfectly intact. Restricting to scalar quantities these are e.\ g.\ correlation functions of products of curvature invariants. Inserting the split \pref{split} into these correlation functions yields a polynomial in $\gamma$ and $\gamma^{-1}$. Using \pref{dyson} allows then to rephrase it as a polynomial only in $\gamma$ to a certain order, which then can be calculated in a consistent perturbative order in $\gamma$ using the usual machinery of perturbation theory. This augmented perturbation theory has the advantage that it maintains automatically diffeomorphism invariance up to the order evaluated, just as in flat space-time \cite{Maas:2020kda}.

\section{First results}

One of the subtleties in quantum gravity is that arguments do not work as in flat-space-time QFT \cite{Maas:2019eux,Schaden:2015wia,Ambjorn:2012jv}. The reason is that in flat space-time these are coordinates, which themselves become (gauge-dependent) quantities in quantum gravity. The only geometric notion of arguments of fields are events. The distance between events is thus itself a correlation function, which must be suitably defined. A possibility is \cite{Maas:2019eux,Schaden:2015wia,Ambjorn:2012jv}
\be
r(X,Y)=\la \min_{z(t)}\int_{X}^{Y} dt g^\mn \frac{dz_\mu(t)}{dt}\frac{dz_\nu(t)}{dt}\ra.\label{diffdist}
\ee
\no In this the minimization over the path $z(t)$ connecting the events $x$ and $y$ should state to find the geodesic length. In leading order in $g^c$, i.\ e.\ $g=g^c$, this automatically yields the geodesic length $r^c$ in the metric $g^c$, and thus what would be expected for QFT in a fixed metric \cite{Maas:2019eux}. Especially, for $g^c$ being the Minkowski metric this yields automatically the correct distance for flat-space-time QFT.

Note that this prevents Fourier space to work as usual. Especially, a usual Fourier transformation will become gauge-dependent. Hence, structures in Fourier space do not necessarily correspond any longer to physical information.

Correlation functions are then given as expectation values of the field operators
\be
\la O(X_1)...O(X_n)\ra=C\left(\left\{r(X_i,X_j)\right\}\right)\nn.
\ee
\no The simplest, non-trivial correlation functions are propagators. Treating them within the FMS approach becomes a double expansion. One is the usual expansion in the coupling constants, and one is the expansion in $g^c$ using \pref{split} and \pref{dyson}, which also needs to be performed in the gauge-fixed path integral before expanding in the other couplings.

At lowest orders in both expansions, the propagator of a scalar, real particle $\phi$ becomes \cite{Maas:2019eux}
\be
\la\phi(X)\phi(Y)\ra\approx D_\text{tl}(r^c(X,Y))+\op(\text{higher order})\label{sprop}
\ee
\no where $D_\text{tl}$ is the tree-level propagator in the metric $g^c$, and thus flat-space-time physics is recovered.

Likewise, if the operator $A$ itself contains the metric, also the operator needs to be expanded, using both \pref{split} and \pref{dyson}. If only the metric itself is involved, this will yield a finite polynomial of correlation functions in $\gamma$, starting with an expectation value containing only $g^c$. If the inverse metric is involved, the expression becomes an infinite series, but still contains only the inverse of $g^c$ in the first term \cite{Maas:2019eux},
\be
\la A(g,g^{-1}) \ra\stackrel{\pref{split}}{=}\la A(g^c,g^{c})^{-1}\ra+\sum_{i=1}^{n<\infty}\op(\gamma^i,(g^c)^{-1})+\sum_{\stackrel{j=1}{i=0}}^{\stackrel{m<\infty}{n<\infty}}\op(\gamma^i,(\gamma^{-1})^j)\stackrel{\pref{dyson}}{=}\la A(g^c,g^{c})^{-1}\ra+\sum_{i=0}^\infty\op(\gamma^i)\nn
\ee
\no Provided the first term is dominant, quantum gravity fluctuations become irrelevant. However, it alone is not diffeomorphism invariant, only the sum of all terms is. In principle, one could also determine the individual terms non-perturbatively. However, only if a perturbative expansion in the coupling and only a few of the additional terms contribute quantitatively relevant, the FMS mechanism really yields an advantage compared to a full non-perturbative calculation.

A particular challenge are those cases in which the leading term is constant, and the dynamical information resides in higher orders. Typical examples are those involving only the metric, e.\ g.\ the scalar geon propagator \cite{Wheeler:1955zz,Maas:2019eux,Maas:2022lxv}, being a correlation function of two curvature scalars. Expanding it to the first non-trivial order yields \cite{Maas:2022lxv}
\be
\la R(X) R(Y)\ra=\left( \pd^{c\mu} \pd^{c\nu} - (g^c)^{\mu \nu} (\pd^c_X)^2 - R^{\mu \nu}_c\right) \left( \pd^{c\rho} \pd^{c\sigma} - (g^c)^{\rho \sigma} (\pd^2_Y)^2 -R^{\rho \sigma}_c \right)\langle \gamma_{\mu \nu}(X) \gamma_{\rho \sigma}(Y)\rangle\nn,
\ee
\no where indices have been raised to this order with the classical metric, and $R_\mn$ is the Ricci tensor. Thus, the scalar geon propagator is given by the contracted elementary graviton propagator. This duality between bound states and elementary particles is typical for the FMS mechanism \cite{Maas:2017wzi}.

Inserting for $g^c$ the flat Minkowski metric and the tree-level graviton propagator yields \cite{Maas:2022lxv}
\be
\la R(X) R(Y)\ra=-i 6 \kappa^2\pd^{c2}_Y\delta^{(4)} \left(x(X) - y(Y)\right)\nn,
\ee
\no where $x(X)$ denote the corresponding coordinates created from $g^c$. Thus, at this lowest order at zero cosmological constant there is no propagating scalar geon. An extension to other $g^c$ (and thus non-zero cosmological constants) or higher order is a formidable endeavor indeed.

\section{Testing the FMS mechanism using dynamical triangulation}

In flat space-time it was very successful to compare the results of the FMS mechanism to non-perturbative lattice results to establish its validity \cite{Maas:2017wzi}. Doing so also for quantum gravity would be desirable, especially as the effort needed for calculations is higher than in flat-space-time theories. The complication arising is, of course, to have a valid quantum gravity lattice version of the same action. Dynamical triangulation \cite{Loll:2019rdj,Ambjorn:2022naa,Ambjorn:2012jv} appears here a suitable candidate, given its compatibility with asymptotic safety results \cite{Reuter:2019byg,Bonanno:2020bil}, which are based also on an Einstein-Hilbert Lagrangian and using the metric as integration variable.

There are now two possibilities to test the FMS mechanism. The first is to determine results using the FMS mechanism for diffeomorphism-invariant operators, and compare them to results from dynamical triangulation. The fact that classical space-times, especially de Sitter space-time, is recovered in dynamical triangulation \cite{Loll:2019rdj,Ambjorn:2022naa,Ambjorn:2012jv}, is already encouraging for the underlying assumption of the FMS mechanism that a classical metric is a good expansion point. Consequently, the result \pref{sprop} has been found to be in agreement with (Euclidean) dynamical triangulation calculations \cite{Dai:2021fqb}.

Another possibility is to determine whether the split \pref{split} is justified, by testing whether $\gamma\ll g^c\approx g$ is indeed true by measuring the fluctuations of $\gamma$ directly, as was done in flat-space-time QFT \cite{Maas:2012ct}. However, this is complicated by the fact that dynamical triangulation, in contrast to flat-space-time lattice gauge theory, does not explicitly manipulate the metric, but creates configurations of invariant lengths $r(X,Y)$, being geodesics between a discrete set of events. It is thus necessary to reconstruct the metric from this information. As the metric is diffeomorphism-variant, this requires to fix a gauge, here the Haywood gauge.

In the discrete set of events this amounts to solving a set of equations,
\bea
r(X_i,X_{i+e})&=&e^\mu g(X_i)_{\mu\rho} e^\rho\nn\\
0&=&g^{\mn}(X_i)\left(g_{\nu\rho}(X_i)-g_{\nu\rho}(X_{i+e})\right)/\Delta(e)\nn
\eea
\no for the symmetric metric. Herein is $e$ a coordinate vector linking the two events, $\Delta(e)$ the corresponding coordinate length, and $r$ being either the constant values $a>0$ or $b<0$, which correspond to the fixed time-like distance and space-like distance between neighboring events in the triangulation \cite{Ambjorn:2012jv}. Information about neighboring events can always be reduced to these equations. This is a finite system of fourth-order polynomial equations. The existence of a solution is guaranteed by construction, though it may not be unique, if the Haywood gauge should turn out to have a Gribov-Singer problem \cite{Maas:2017wzi,Maas:2019eux}. Finding an efficient algorithm to solve the system, and thus reconstruct the metric, is an ongoing project.

\section{Summary}

The observation of a large-scale fixed metric suggests that quantum gravity should be treatable as perturbations around it. The FMS mechanism provides a possibility to do so and maintain diffeomorphism invariance of observables systematically, and especially beyond perturbation theory. In contrast to flat space-time QFT \cite{Maas:2017wzi}, this requires a second infinite series expansion besides the usual infinite perturbative expansion \cite{Maas:2022lxv}. While there is still a systematic ordering of the series, it becomes much more challenging in practice. Thus, testing the FMS mechanism in non-perturbative numerical simulations would make the case to invest the effort better. This appears possible from multiple directions using dynamical triangulation, and thus is worthwhile to explore.

\bibliographystyle{bibstyle}
\bibliography{bib}

\begin{thebibliography}{10}

\bibitem{Frohlich:1980gj}
J.~Fr\"ohlich, G.~Morchio, and F.~Strocchi,
\newblock Phys.Lett. {\bf B97}, 249 (1980).
%%CITATION = PHLTA,B97,249;%%

\bibitem{Maas:2017wzi}
A.~Maas,
\newblock Progress in Particle and Nuclear Physics {\bf 106}, 132 (2019),
  1712.04721.
%%CITATION = ARXIV:1712.04721;%%

\bibitem{Maas:2019eux}
A.~Maas,
\newblock SciPost Phys. {\bf 8}, 051 (2020), 1908.02140.
%%CITATION = ARXIV:1908.02140;%%

\bibitem{Reuter:2019byg}
M.~Reuter and F.~Saueressig,
\newblock {\em {Quantum Gravity and the Functional Renormalization Group}}
  (Cambridge University Press, 2019).
%%CITATION = INSPIRE-1716753;%%

\bibitem{Bonanno:2020bil}
A.~Bonanno {\em et~al.},
\newblock Front. in Phys. {\bf 8}, 269 (2020), 2004.06810.

\bibitem{Loll:2019rdj}
R.~Loll,
\newblock Class. Quant. Grav. {\bf 37}, 013002 (2020), 1905.08669.
%%CITATION = ARXIV:1905.08669;%%

\bibitem{Ambjorn:2022naa}
J.~Ambjorn,
\newblock {Lattice Quantum Gravity: EDT and CDT},
\newblock 2022, 2209.06555.

\bibitem{Hehl:1976kj}
F.~W. Hehl, P.~Von Der~Heyde, G.~D. Kerlick, and J.~M. Nester,
\newblock Rev. Mod. Phys. {\bf 48}, 393 (1976).
%%CITATION = RMPHA,48,393;%%

\bibitem{Maas:2022lxv}
A.~Maas, M.~Markl, and M.~M\"uller,
\newblock (2022), 2202.05117.

\bibitem{Frohlich:1981yi}
J.~Fr\"ohlich, G.~Morchio, and F.~Strocchi,
\newblock Nucl.Phys. {\bf B190}, 553 (1981).
%%CITATION = NUPHA,B190,553;%%

\bibitem{Maas:2020kda}
A.~Maas and R.~Sondenheimer,
\newblock Phys. Rev. D {\bf 102}, 113001 (2020), 2009.06671.

\bibitem{Schaden:2015wia}
M.~Schaden,
\newblock (2015), 1509.03095.
%%CITATION = ARXIV:1509.03095;%%

\bibitem{Ambjorn:2012jv}
J.~Ambjorn, A.~Goerlich, J.~Jurkiewicz, and R.~Loll,
\newblock Phys. Rept. {\bf 519}, 127 (2012), 1203.3591.
%%CITATION = ARXIV:1203.3591;%%

\bibitem{Wheeler:1955zz}
J.~A. Wheeler,
\newblock Phys. Rev. {\bf 97}, 511 (1955).
%%CITATION = PHRVA,97,511;%%

\bibitem{Dai:2021fqb}
M.~Dai, J.~Laiho, M.~Schiffer, and J.~Unmuth-Yockey,
\newblock Phys. Rev. D {\bf 103}, 114511 (2021), 2102.04492.

\bibitem{Maas:2012ct}
A.~Maas,
\newblock Mod. Phys. Lett. {\bf A27}, 1250222 (2012), 1205.0890.

\end{thebibliography}

\end{document}